\documentclass[a4paper,10pt]{article}
\usepackage{stmaryrd}
\usepackage{amsfonts}
\usepackage{bbm}
\usepackage{amscd}
\usepackage{mathrsfs}
\usepackage{latexsym,amssymb,amsmath,amscd,amscd,amsthm,amsxtra,xypic}
\usepackage[dvips]{graphicx}
\usepackage[utf8]{inputenc}
\usepackage[T1]{fontenc}
\usepackage{lmodern}
\usepackage{amssymb}
\usepackage[all]{xy}
\usepackage{nicefrac,mathtools,enumitem}
\usepackage{microtype}

\newtheorem{thm}{Theorem}[section]
\newtheorem{lem}[thm]{Lemma}
\newtheorem{cor}[thm]{Corollary}
\newtheorem{pro}[thm]{Proposition}
\newtheorem{ex}[thm]{Example}
\newtheorem{rmk}[thm]{Remark}
\newtheorem{defi}[thm]{Definition}
\newtheorem{pdef}[thm]{Proposition-Definition}

\newcommand{\be }{\begin{equation}}
\newcommand{\ee }{\end{equation}}

\newcommand{\pf}{\noindent{\bf Proof.}\ }

\newcommand{\huaV}{\mathcal{V}}

\newcommand{\frkg}{\mathfrak g}
\newcommand{\frkh}{\mathfrak h}

\newcommand{\frkk}{\mathfrak k}

\def\qed{\hfill ~\vrule height6pt width6pt depth0pt}

\newcommand{\br}[1]{   [ \cdot,    \cdot  ]   }

\newcommand{\Id}{\rm{Id}}

\newcommand{\dM}{\mathrm{d}}

\newcommand{\Hom}{\mathrm{Hom}}

\newcommand{\tr}{\mathrm{tr}}

\newcommand{\gl}{\mathfrak {gl}}
\newcommand{\sln}{\mathfrak {sl}}
\newcommand{\so}{\mathfrak {so}}

\newcommand{\Ker}{\mathrm{Ker}}

\newcommand{\End}{\mathrm{End}}
\newcommand{\ad}{\mathrm{ad}}

\newcommand{\ve}{\mathrm{v}}
\newcommand{\Vect}{\mathrm{Vect}}
\newcommand{\sgn}{\mathrm{sgn}}
\newcommand{\Ksgn}{\mathrm{Ksgn}}

\newcommand{\V}{\mathbb{V}}

\begin{document}
\title{
{Hom-Lie 2-algebras
\thanks
 {
Research partially supported by NSFC  (11101179) and SRFDP
(20100061120096).
 }
} }
\author{
Yunhe Sheng, Danhua Chen  \\
Department of Mathematics, Jilin University,\\
 Changchun 130012, Jilin, China
\\\vspace{3mm}
email: shengyh@jlu.edu.cn }

\date{}


\maketitle
\begin{abstract}
In this paper, we introduce the notions of hom-Lie 2-algebras, which
is the categorification of hom-Lie algebras, $HL_\infty$-algebras,
which is the hom-analogue of $L_\infty$-algebras, and crossed
modules of hom-Lie algebras. We prove that the category of hom-Lie
2-algebras and the category of 2-term $HL_\infty$-algebras are
equivalent. We give a detailed study on skeletal hom-Lie 2-algebras.
In particular, we construct the hom-analogues of the string Lie
2-algebras associated to any semisimple involutive hom-Lie algebras.
We also proved that there is a one-to-one correspondence between
strict hom-Lie 2-algebras and crossed modules of hom-Lie algebras.
We give the construction of strict hom-Lie 2-algebras from
hom-left-symmetric algebras and symplectic hom-Lie algebras.
\end{abstract}

\tableofcontents
\section{Introduction}

The notion of hom-Lie algebras was introduced by Hartwig, Larsson,
and Silvestrov in \cite{HLS} as part of a study of deformations of
the Witt and the Virasoro algebras. In a hom-Lie algebra, the Jacobi
identity is twisted by a linear map, called the hom-Jacobi identity.
Some $q$-deformations of the Witt and the Virasoro algebras have the
structure of a hom-Lie algebra \cite{HLS}. Because of close relation
to discrete and deformed vector fields and differential calculus
\cite{HLS,LD1,LD2}, hom-Lie algebras are widely studied recently
\cite{AMM,BM,MS1,MS2,shenghomLie,Yao1,Yao2,Yao3,BaiHom}.

Recently, people have payed more attention to higher categorical
structures with motivations from string theory
\cite{baez:classicalstring}. One way to provide higher categorical
structures is by categorifying existing mathematical concepts. One
of the simplest higher structure is a $2$-vector space, which is a
categorified  vector space. If we further put Lie algebra structures
on $2$-vector spaces, then we obtain the notion of Lie $2$-algebras
\cite{baez:2algebras}. The Jacobi identity is replaced by a natural
transformation, called Jacobiator, which also satisfies some
coherence laws of its own. One of the motivating examples is the
differentiation of Witten's string Lie 2-group $String(n)$, which is
called a string Lie $2$-algebra. As $SO(n)$ is the connected part of
$O(n)$ and $Spin(n)$ is the simply connected cover of $SO(n)$,
 $String(n)$ is a ``cover'' of $Spin(n)$ which has trivial $\pi_3$ (notice that $\pi_2(G)=0$ for any Lie group $G$). The differentiation of $String(n)$ is not
 any more  $\so (n)$, but a central extension of $\so (n)$ by the abelian Lie $2$-algebra  $\mathbb R \to 0$, which is a Lie $2$-algebra by
  itself. The concept of string Lie $2$-algebra is later generalized to any such extension of a semisimple Lie
  algebra. $L_\infty$-algebras,  sometimes  called strongly
homotopy (sh) Lie algebras,  were introduced \cite{Stasheff1} as a
model for ``Lie algebras that satisfy Jacobi identity up to all
higher homotopies''. It is well known that Lie 2-algebras are
equivalent to 2-term $L_\infty$-algebras.

In this paper, we provide the categorification of hom-Lie algebras,
which we call hom-Lie 2-algebras. We also give the hom-analogue of
$L_\infty$-algebras, which we call $HL_\infty$-algebras. The main
difficulty to give these definitions is how to let the
hom-structures involved in. In the case of Lie 2-algebras  (or
2-term $L_\infty$-algebras), the Jacobiator (or $l_3$) should
satisfy some kind of closed condition. Motivated by the cohomology
theory introduced in \cite{shenghomLie}, we solve this difficulty
successfully. We prove that the category of hom-Lie 2-algebras and
the category of 2-term $HL_\infty$-algebras are equivalent. Skeletal
hom-Lie 2-algebras are studied in detail. We give their
classification by the third cohomology of hom-Lie algebras, and
provide examples from quadratic hom-Lie algebras introduced in
\cite{BM} by Benayadi and Makhlouf. In particular, we introduce the
hom-analogues of the string Lie 2-algebras. The notion of crossed
modules of hom-Lie algebras is also introduced and we prove that
there is a one-to-one correspondence between strict hom-Lie
2-algebras and crossed modules of hom-Lie algebras. We construct
strict hom-Lie 2-algebras from hom-left-symmetric algebras.
Furthermore, we introduce the notion of a symplectic hom-Lie
algebra, which is a hom-Lie algebra together with a symplectic form,
and give the construction of strict hom-Lie 2-algebras from
symplectic hom-Lie algebras.

The paper is organized as follows. In Section 2, we recall some
necessary background knowledge, including the cohomology theory of
hom-Lie algebras and 2-vector spaces. In Section 3, first we give
the definition of hom-Lie 2-algebras, which is the categorification
of hom-Lie algebras. Then we introduce the hom-analogue of
$L_\infty$-algebras, which we call $HL_\infty$-algebras. We give the
definition of 2-term $HL_\infty$-algebras by explicit formulas. At
last, we prove that the category of hom-Lie 2-algebras and the
category of 2-term $HL_\infty$-algebras are equivalent. In Section
4, we study skeletal hom-Lie 2-algebras. Especially, we construct
examples of skeletal hom-Lie 2-algebras from involutive quadratic
hom-Lie algebras, and obtain hom-analogues of string Lie 2-algebras.
In Section 5, first we introduce the notion of crossed modules of
hom-Lie algebras, and prove that they are equivalent to strict
hom-Lie 2-algebras. We construct strict hom-Lie 2-algebras from
hom-left-symmetric algebras. At last, we introduce the notion of
symplectic hom-Lie algebras. There is a natural hom-left-symmetric
algebra associated to a symplectic hom-Lie algebra, such that it is
the sub-adjacent hom-Lie algebra of the induced hom-left-symmetric
algebra. Then we give the construction of strict hom-Lie 2-algebras
from symplectic hom-Lie algebras.

{\bf Acknowledgement:} We give our warmest thanks to the referee for
very helpful comments.

\section{Preliminaries}

In this section, we recall some basic notions and facts about
hom-Lie algebras \cite{HLS} and 2-vector spaces
\cite{baez:2algebras}.\vspace{1mm}

 \noindent $\bullet$ {\bf Hom-Lie algebras and their
representations}

\begin{defi}{\rm\cite{HLS}}
  A hom-Lie algebra is a triple $(\frkg,\br__\frkg ,\phi_\frkg)$ consisting of a
  linear space $\frkg$, a skew-symmetric bilinear map (bracket) $\br,_\frkg:\wedge^2\frkg\longrightarrow
  \frkg$ and an algebra morphism $\phi_\frkg:\frkg\longrightarrow\frkg$ satisfying
  \begin{equation}
    [\phi_\frkg(u),[v,w]_\frkg]_\frkg+[\phi_\frkg(v),[w,u]_\frkg]_\frkg+[\phi_\frkg(w),[u,v]_\frkg]_\frkg=0.
  \end{equation}
The hom-Lie algebra $(\frkg,\br__\frkg ,\phi_\frkg)$ is said to be
regular (involutive), if $\phi_\frkg$ is nondegenerate (satisfies
$\phi_\frkg^2=\Id$);
\end{defi}

\begin{rmk}
  There is a more general notion of hom-Lie algebras introduced by Makhlouf and
  Silvestrov in \cite{MS2}, in which $\phi_\frkg$ is only a
  homomorphism of linear spaces. A hom-Lie algebra in this paper is
   called a multiplicative hom-Lie algebra in \cite{BM}
\end{rmk}

\begin{defi} A morphism of hom-Lie algebras $f:(\frkg,\br__\frkg
,,\phi_\frkg)\longrightarrow(\frkk,\br__\frkk,\phi_\frkk)$ is a
linear map $f:\frkg\longrightarrow\frkk$ such that
\begin{eqnarray}
\label{eqn:phimorphism1}f[u,v]_\frkg&=&[f(u),f(v)]_\frkk,\quad\forall~u,v\in\frkg,\\\label{eqn:phimorphism2}
f\circ\phi_\frkg&=&\phi_\frkk\circ f.
\end{eqnarray}
\end{defi}

Let $(\frkg,\br__\frkg,\phi_\frkg)$ be a hom-Lie algebra and $V$ an
arbitrary vector space. Let $A\in\gl(V)$ be an arbitrary linear
transformation from $V$ to $V$. The representation of hom-Lie
algebras was introduced in \cite{shenghomLie}.

\begin{defi}\label{defi:representation}
  A representation of the hom-Lie algebra $(\frkg,\br,_\frkg,\phi_\frkg)$ on
  the vector space $V$ with respect to $A\in\gl(V)$ is a linear map
  $\rho_A:\frkg\longrightarrow \gl(V)$, such that for any
  $u,v\in\frkg$, the following equalities are satisfied:
  \begin{itemize}
\item[\rm(i)] $\rho_A(\phi_\frkg(u))\circ A=A\circ
\rho_A(u);$
\item[\rm(ii)] $
    \rho_A([u,v]_\frkg)\circ
    A=\rho_A(\phi_\frkg(u))\circ\rho_A(v)-\rho_A(\phi_\frkg(v))\circ\rho_A(u).$
   \end{itemize}
\end{defi}
The set of $k$-cochains on $\frkg$ with values in $V$, which we
denote by $C^k(\frkg;V)$, is the set of  skew-symmetric $k$-linear
maps from $\frkg\times\cdots\times\frkg$ $(k$-times$)$ to $V$:
$$
C^k(\frkg;V)\triangleq\{f:\wedge^k\frkg\longrightarrow V ~\mbox{is a
$k$-linear map}\}.
$$

A $k$-hom-cochain on $\frkg$ with values in $V$ is defined to be a
$k$-cochain $f\in C^k(\frkg;V)$ such that it is compatible with
$\phi_\frkg$ and $A$ in the sense that $A\circ f=f\circ \phi_\frkg$,
i.e.
$$
A(f(u_1,\cdots,u_k))=f(\phi_\frkg(u_1),\cdots,\phi_\frkg(u_k)).
$$
Denote by $ C^k_{\phi_\frkg,A}(\frkg;V)$ the set of
$k$-hom-cochains:
$$
C^k_{\phi_\frkg,A}(\frkg;V)\triangleq\{f\in C^k(\frkg;V)|~A\circ
f=f\circ \phi_\frkg\}.
$$

In \cite{shenghomLie}, the author defined the coboundary operator
$\dM_{\rho_A}:C^k_{\phi_\frkg,A}(\frkg;V)\longrightarrow
C^{k+1}_{\phi_\frkg,A}(\frkg;V)$ by setting
\begin{eqnarray}
\nonumber  \dM_{\rho_A} f(u_1,\cdots,u_{k+1})&=&\sum_{i=1}^{k+1}(-1)^{i+1}\rho(\phi_\frkg^{k-1}(u_i))(f(u_1,\cdots,\widehat{u_i},\cdots,u_{k+1}))\\
 \label{eqn:d} &&+\sum_{i<j}(-1)^{i+j}f([u_i,u_j]_\frkg,\phi_\frkg(u_1)\cdots,\widehat{u_i},\cdots,\widehat{u_j},\cdots,\phi_\frkg(u_{k+1})).
\end{eqnarray}
The equality $\dM_{\rho_A}^2=0$ was proved in \cite{shenghomLie}.
Thus, we can obtain the cohomology of hom-Lie algebras.

\begin{rmk}
  The formula given by \eqref{eqn:d} is  slightly different from the
  formula given in \cite{shenghomLie}, where the author use
  $\rho(\phi_\frkg^{k}(u_i))$. Both of them are correct. All the
  results in \cite{shenghomLie} hold after a small modification for the coboundary
  operator $\dM_{\rho_A}$ given by \eqref{eqn:d}.
\end{rmk}

Every hom-Lie algebra has the trivial representation on $\mathbb R$
with respect to $\Id:\mathbb R\longrightarrow \mathbb R$, the
corresponding coboundary operator, which we denote by $\dM_T$, is
given by
\begin{eqnarray*}
\dM_T
f(u_1,\cdots,u_{k+1})&=&\sum_{i<j}(-1)^{i+j}f([u_i,u_j]_\frkg,\phi_\frkg(u_1)\cdots,\widehat{u_i},\cdots,\widehat{u_j},\cdots,\phi_\frkg(u_{k+1})).
\end{eqnarray*}
Denote by $Z^k_{\phi_\frkg}(\frkg)$ and $B^k_{\phi_\frkg}(\frkg)$
the corresponding closed $k$-hom-cochains and exact $k$-hom-cochains
respectively. Denote the resulting cohomology by
$H^k(\frkg)$.\vspace{2mm}

\noindent $\bullet$ {\bf 2-vector spaces}\vspace{2mm}

Vector spaces can be categorified to $2$-vector spaces. A good
introduction for this subject is \cite{baez:2algebras}. Let $\Vect$
be the category of vector spaces.

\begin{defi}{\rm\cite{baez:2algebras}}
A $2$-vector space is a category in the category $\Vect$.
\end{defi}

Thus, a $2$-vector space $C$ is a category with a vector space of
objects $C_0$ and a vector space of morphisms $C_1$, such that all
the structure maps are linear. Let $i:C_0\longrightarrow C_1$ be the
identity assigning map and $s,t:C_1\longrightarrow C_0$ be the
source and target maps respectively. Let $\cdot_\ve$ be the
composition of morphisms.

It is well known that the 2-category of $2$-vector spaces is
equivalent to the 2-category of 2-term complexes of vector spaces.
Roughly speaking, given a $2$-vector space $C$,
\begin{equation}\label{eqn:2term}
\Ker(s)\stackrel{t}{\longrightarrow}C_0
\end{equation} is a 2-term complex.
Conversely, any 2-term complex of vector spaces
$V_1\stackrel{\dM}{\longrightarrow}V_0$ gives rise to a $2$-vector
space of which the set of objects is $V_0$, the set of morphisms is
$V_0\oplus V_1$, the source map $s$ is given by $s(v,m)=v$, and the
target map $t$ is given by $t(v,m)=v+\dM m$, where $v\in V_0,~m\in
V_1.$ We denote the $2$-vector space associated to the 2-term
complex of vector spaces $V_1\stackrel{\dM}{\longrightarrow}V_0$ by
$\mathbb{V}$:
\begin{equation}\label{v}
\mathbb{V}=\begin{array}{c}
 \mathbb{V}_1:=V_0\oplus V_1\\
\vcenter{\rlap{s }}~\Big\downarrow\Big\downarrow\vcenter{\rlap{t }}\\
 \mathbb{V}_0:=V_0.
 \end{array}\end{equation}

Given a $2$-vector space $\V$, we define $\End^0_\dM(\V)$ by
$$
\End^0_\dM(\V)\triangleq\{(A_0,A_1)\in\gl(V_0)\oplus
\gl(V_1)|A_0\circ\dM=\dM\circ A_1\},
$$
and define $\End^1(\V)\triangleq \Hom(V_0,V_1)$. Then we have,
\begin{lem}{\rm\cite{SLZ}} $\End^0_\dM(\V)$ is the space of linear functors from
  $\mathbb{V}$ to   $\mathbb{V}$.
 \end{lem}
There is a differential $\delta:\End^1(\V)\longrightarrow
\End^0_\dM(\V)$ given by
$$
\delta(\alpha)\triangleq(\dM\circ\alpha,\alpha\circ\dM),\quad\forall~\alpha\in\End^1(\V),
$$
and a bracket operation $[\cdot,\cdot]_C$ given by the graded
commutator. More precisely,  for any $A=(A_0,A_1),B=(B_0,B_1)\in
\End^0_\dM(\V)$ and $\alpha\in\End^1(\V)$, $[\cdot,\cdot]_C$ is
given by
\begin{eqnarray*}
  [A,B]_C=A\circ B-B\circ A=(A_0\circ B_0-B_0\circ A_0,A_1\circ B_1-B_1\circ
  A_1),
\end{eqnarray*}
and
\begin{equation}\label{representation}
  ~[A,\alpha]_C=A\circ \alpha-\alpha\circ A=A_1\circ \alpha-\alpha\circ A_0.
\end{equation}
These two operations make $\End^1(\V)\xrightarrow{\delta}
\End^0_\dM(\V)$ into a 2-term DGLA (proved in \cite{shengzhu2}),
which we denote by $\End(\V)$. It plays the same role  as $\gl(V)$
for a vector space $V$ in the classical case.

\section{Hom-Lie 2-algebras and $HL_\infty$-algebras}
In this section, first we category the notion of hom-Lie algebras,
and obtain the hom-Lie 2-algebras. Then we give the hom-analogue of
$L_\infty$-algebras, what we call $HL_\infty$-algebras. We give the
structure of a 2-term $HL_\infty$-algebra by explicit formulas.
 At last, we prove that the category of hom-Lie
2-algebras and the category of 2-term $HL_\infty$-algebras are
equivalent.

\subsection{Hom-Lie 2-algebras}

\begin{defi}\label{defi:Lie2}
A hom-Lie 2-algebra is a 2-vector space $L$ equipped with
\begin{itemize}
\item[$\bullet$] a skew-symmetric bilinear functor, the bracket, $[\cdot,\cdot]:L\times L\longrightarrow
L$,
\item[$\bullet$] a linear functor $\Phi=(\Phi_0,\Phi_1):L\longrightarrow L$
satisfying:
\begin{equation}\label{eqn:Phimorphism}
\Phi[\xi,\eta]=[\Phi(\xi),\Phi(\eta)],\quad\forall~\xi,\eta\in L.
\end{equation}
\item[$\bullet$] a skew-symmetric trilinear natural isomorphism, the
hom-Jacobiator,
$$
J_{x,y,z}:[[x,y],\Phi_0(z)]\longrightarrow
[\Phi_0(x),[y,z]]+[[x,z],\Phi_0(y)],
$$
satisfying $J_{\Phi_0(x),\Phi_0(y),\Phi_0(z)}=\Phi_1J_{x,y,z}$,
\end{itemize}
such that the following hom-Jacobiator identity is satisfied,
\begin{eqnarray*}
&&J_{[w,x],\Phi_0(y),\Phi_0(z)}\cdot_\ve([J_{w,x,z},\Phi^2_0(y)]+1)\cdot_\ve(J_{\Phi_0(w),[x,z],\Phi_0(y)}+J_{[w,z],\Phi_0(x),\Phi_0(y)}+1)\\
&=&[J_{w,x,y},\Phi^2_0(z)]\cdot_\ve(J_{[w,y],\Phi_0(x),\Phi_0(z)}+J_{\Phi_0(w),[x,y],\Phi_0(z)})\\
&&\cdot_\ve([\Phi^2_0(w),J_{x,y,z}]+[J_{w,y,z},\Phi^2_0(x)]+1)\cdot_\ve(1+J_{\Phi_0(w),[y,z],\Phi_0(x)}),
\end{eqnarray*}
or, in terms of a diagram,
$$
\footnotesize{\xymatrix{
[[[w,x],\Phi_0(y)],\Phi^2_0(z)]\ar[d]_{[J_{w,x,y},\Phi^2_0(z)]}\ar[rr]^{J_{[w,x],\Phi_0(y),\Phi_0(z)}\qquad\quad}&&[\Phi_0[w,x],[\Phi_0(y),\Phi_0(z)]]
+[[[w,x],\Phi_0(z)],\Phi^2_0(y)]\ar[d]^{1+[J_{w,x,z},\Phi^2_0(y)]}\\
[[\Phi_0(w),[x,y]],\Phi^2_0(z)]+[[[w,y],\Phi_0(x)],\Phi^2_0(z)]\ar[d]^{J_{\Phi_0(w),[x,y],\Phi_0(z)}+J_{[w,y],\Phi_0(x),\Phi_0(z)}}&&
M\ar[d]^{ 1+J_{\Phi_0(w),[x,z],\Phi_0(y)}+J_{[w,z],\Phi_0(x),\Phi_0(y)}}\\
P\ar[rr]_{(1+[\Phi^2_0(w),J_{x,y,z}]+[J_{w,y,z},\Phi^2_0(x)])\cdot_\ve(1+J_{\Phi_0(w),[y,z],\Phi_0(x)})}&&
Q, }}
$$
where $M,P$ and $Q$ are given by
\begin{eqnarray*}
M&=&[\Phi_0[w,x],[\Phi_0(y),\Phi_0(z)]]+[[\Phi_0(w),[x,z]],\Phi^2_0(y)]+[[[w,z],\Phi_0(x)],\Phi^2_0(y)];\\
P&=&[\Phi^2_0(w),[[x,y],\Phi_0(z)]]+[[\Phi_0(w),\Phi_0(z)],\Phi_0[x,y]]\\&&+[\Phi_0[w,y],[\Phi_0(x),\Phi_0(z)]]+[[[w,y],\Phi_0(z)],\Phi^2_0(x)];\\
Q&=&[\Phi_0[w,x],[\Phi_0(y),\Phi_0(z)]]+[\Phi^2_0(w),[[x,z],\Phi_0(y)]]+[[\Phi_0(w),\Phi_0(y)],\Phi_0([x,z])]\\
&&+[\Phi_0([w,z]),[\Phi_0(x),\Phi_0(y)]]+[[[w,z],\Phi_0(y)],\Phi^2_0(x)].
\end{eqnarray*}
\end{defi}

Usually we denote a hom-Lie 2-algebra by $(L,[\cdot,\cdot],J,\Phi)$.
A hom-Lie 2-algebra is called {\em strict} if the hom-Jacobiator is
the identity isomorphism.

\begin{defi}
  Given hom-Lie 2-algebras $(L,\br,, \Phi)$ and $(L^\prime,\br,', \Phi^\prime)$, a hom-Lie 2-algebra morphism $F:L\longrightarrow
  L^\prime$ consists of:
  \begin{itemize}
   \item[$\bullet$] a linear functor $(F_0,F_1)$ from the underlying
   2-vector space of $L$ to that of $L^\prime$ such that
    $$
    \Phi^\prime\circ (F_0,F_1)=(F_0,F_1)\circ\Phi,
    $$
 \item[$\bullet$] a skew-symmetric bilinear natural transformation
 $$
F_2(x,y):[F_0(x),F_0(y)]'\longrightarrow F_0([x,y])
 $$
satisfying $F_2(\Phi_0(x),\Phi_0(y))=\Phi_1'F_2(x,y)$,  such that
the following diagram commutes:
 $$
\footnotesize{ \xymatrix{
[[F_0(x),F_0(y)]',\Phi'(F_0(z))]'\ar[d]_{J_{F_0(x),F_0(y),F_0(z)}}\ar[rr]^{\qquad\qquad\qquad[F_2(x,y),1]'\quad\qquad\quad}&&[F_0[x,y],F_0\Phi(z)]'\ar[d]^{F_2([x,y],\Phi(z))}\\
~[\Phi'(F_0(x)),[F_0(y),F_0(z)]']'+[[F_0(x),F_0(z)]',\Phi'(F_0(y))]'\ar[d]_{[1,F_2(y,z)]'+[F_2(x,z),1]'}&&F_0[[x,y],\Phi(z)]\ar[d]^{F_1J_{x,y,z}}\\
~[F_0(\Phi(x)),F_0[y,z]]'+[F_0[x,z],F_0(\Phi(y))]'\ar[rr]^{\footnotesize{F_2(\Phi(x),[y,z])+F_2([x,z],\Phi(y))}}&&
\footnotesize{F_0[\Phi(x),[y,z]]+F_0[[x,z],\Phi(y)]. }}}
$$
  \end{itemize}
\end{defi}


The identity morphism ${\Id}_L:L\longrightarrow L$ has the identity
functor as its underlying functor, together with an identity natural
transformation as $({\Id}_L)_2$.  Let $L,~L'$ and $L''$ be hom-Lie
2-algebras, the composition of a pair of hom-Lie 2-algebra morphisms
$F:L\longrightarrow L'$ and $G:L'\longrightarrow L''$, which we
denote by $G\circ F$, is given by letting the functor $((G\circ
F)_0,(G\circ F)_1)$ be the usual composition of $(G_0,G_1)$ and
$(F_0,F_1)$, and letting $(G\circ F)_2$ be the following composite:
$$
 \xymatrix{
[G_0\circ F_0(x),G_0\circ F_0(y)]''\ar[dd]_{G_2(F_0(x),F_0(y))}\ar[dr]^{(G\circ F)_2(x,y)}&&\\
&G_0\circ F_0[x,y].&\\
 G_0[F_0(x),F_0(y)]'\ar[ur]_{G_1(F_2(x,y))}&&,}.
$$

It is straightforward to see that
\begin{pro}
There is a category {\bf HLie2} with hom-Lie 2-algebras as
  objects and hom-Lie 2-algebra morphisms as morphisms.
\end{pro}

\subsection{$HL_\infty$-algebras}

\begin{defi}
 A $HL_\infty$-algebra is a graded  vector space $V_\bullet=\oplus_{i=0}^{\infty}V_i$ equipped with
 \begin{itemize}\item a system  $\{l_k | 1 \leq k < \infty  \}
 $ of linear maps $l_k:\wedge^kV_\bullet\longrightarrow V_\bullet$
 with  $deg(l_k)$ = $k-2$, where the exterior powers are interpreted in the
graded sense, i.e. the following relation with Koszul sign ``Ksgn''
is satisfied:
 $$
 l_k (x_{\sigma (1)}, \cdots , x_{\sigma (k)}) = \sgn(\sigma)\Ksgn(\sigma) l_k (x_1 , \cdots ,
 x_k), \quad\forall \sigma\in S_k;
 $$

\item a system $\{ \phi_k |1 \leq k < \infty \}$ of linear maps
$\phi_k : V_k \rightarrow V_k$,
  such that for any $ x_1 \in V_{s_1}, \ldots , x_k \in V_{s_k}$, we
have
 $$
 \phi_{(\sum s_i)+k-2} (l_k(x_1 , \ldots  ,x_k ))= l_k (\phi_{s_1}(x_1) ,\ldots ,\phi_{s_k}(x_k)),
 $$
  \end{itemize}
  such that the following generalized form of the Jacobi identity holds for all $0 \leq n < \infty$,
 $$
 \sum \limits _{i+j = n+1} \sum \limits_ {\sigma} (-1)^{i(j-1)} \sgn(\sigma)\Ksgn(\sigma) l_j(l_i(x_{\sigma(1)},\cdots, x_{\sigma(i)} ),
 \phi_{m_{i+1}}^{i-1}(x_{\sigma (i+1)}), \cdots , \phi_{m_{n}}^{i-1}(x_{\sigma (n)}) = 0,
 $$
 where $x_{\sigma (i+1)} \in V_{m_{i+1}} , \cdots , x_{\sigma (n)} \in
 V_{m_n}$, and the summation is taken over all $(i,n-i)$-unshuffles with
$i\geq1$.
\end{defi}
\begin{rmk}
  A similar notion of $HL_\infty$-algebras was given by Yau in
  \cite{Yau4}. We will see that our definition fits very well with
  the cohomology theory of hom-Lie algebras introduced in
  \cite{shenghomLie}.
\end{rmk}

For $n=1$, we have
$$
l_1^2=0,\quad l_1:V_{i+1}\longrightarrow V_i,
$$
which means that $V_\bullet$ is a complex of vector spaces, so  we
write $\dM=l_1$ as usual. For $n=2$, we have
$$
\dM l_2(x_p,x_q)=l_2(\dM x_p,x_q)+(-1)^pl_2(x_p,\dM x_q),\quad
\forall~x_p\in V_p, x_q\in V_q,
$$
which means that $\dM $ is a derivation with respect to $l_2$.

Constraint on the 2-term case, it is not hard to obtain:

\begin{pdef}\label{defi:2hl}
  A 2-term $HL_\infty$-algebra $\huaV$ consists of the following data:
\begin{itemize}
\item[$\bullet$] a complex of vector spaces $V_1\stackrel{\dM}{\longrightarrow}V_0,$

\item[$\bullet$] bilinear maps $l_2,:V_i\times V_j\longrightarrow
V_{i+j}$,

\item[$\bullet$] two linear transformations $\phi_0\in\gl(V_0)$ and $\phi_1\in\gl(V_1)$ satisfying $\phi_0\circ\dM=\dM\circ\phi_1$,
\item[$\bullet$] a skew-symmetric trilinear map $l_3:V_0\times V_0\times V_0\longrightarrow
V_1$ satisfying $l_3\circ\phi_0=\phi_1\circ l_3,$
   \end{itemize}
   such that for any $w,x,y,z\in V_0$ and $m,n\in V_1$, the following equalities are satisfied:
\begin{itemize}
\item[$\rm(a)$] $l_2(x,y)=-l_2(y,x),$
\item[$\rm(b)$] $l_2(x,m)=-l_2(m,x),$
\item[$\rm(c)$] $l_2(m,n)=0,$
\item[$\rm(d)$] $\dM l_2(x,m)=l_2(x,\dM m),$
\item[$\rm(e)$] $l_2(\dM m,n)=l_2(m,\dM n),$
\item[$\rm(f)$] $\phi_0(l_2(x,y))=l_2(\phi_0(x),\phi_0(y)),$
\item[$\rm(g)$]$\phi_1(l_2(x,m))=l_2(\phi_0(x),\phi_1(m)),$
\item[$\rm(h)$]$\dM l_3(x,y,z)=l_2(\phi_0(x),l_2(y,z))+l_2(\phi_0(y),l_2(z,x))+l_2(\phi_0(z),l_2(x,y)),$
\item[$\rm(i)$] $\dM l_3(x,y,\dM m)=l_2(\phi_0(x),l_2(y,m))+l_2(\phi_0(y),l_2(m,x))+l_2(\phi_1(m),l_2(x,y)),$
\item[$\rm(j)$] \begin{eqnarray*}
&&l_3(l_2(w,x),\phi_0(y),\phi_0(z))+l_2(l_3(w,x,z),\phi^2_0(y))\\
&&+
l_3(\phi_0(w),l_2(x,z),\phi_0(y))+l_3(l_2(w,z),\phi_0(x),\phi_0(y)) \\
&=&l_2(l_3(w,x,y),\phi^2_0(z))+l_3(l_2(w,y),\phi_0(x),\phi_0(z))+l_3(\phi_0(w),l_2(x,y),\phi_0(z))\\
&&+l_2(\phi^2_0(w),l_3(x,y,z))+l_2(l_3(w,y,z),\phi^2_0(x))+l_3(\phi_0(w),l_2(y,z),\phi_0(x)).
\end{eqnarray*}
   \end{itemize}
\end{pdef}

We will denote a 2-term $HL_\infty$-algebra by
$(V_1\stackrel{\dM}{\longrightarrow}V_0,l_2,l_3,\phi_0,\phi_1)$.

\begin{defi}\label{defi:morphism}
  Let $\huaV$ and $\huaV^\prime $ be 2-term $HL_\infty$-algebras. A
  $HL_\infty$-morphism $f:\huaV\longrightarrow \huaV^\prime$ consists of:
  \begin{itemize}
    \item[$\bullet$] a chain map $f:\huaV\longrightarrow \huaV^\prime$, which
    consists of linear maps $f_0:V_0\longrightarrow V_0'$ and $f_1:V_1\longrightarrow
    V_1'$ satisfying $$f_0\circ\dM=\dM^\prime\circ f_1,$$ and
\begin{equation}\label{eqn:morphismphi1}
f_0\circ \phi_0=\phi_0^\prime\circ f_0,\quad f_1\circ
\phi_1=\phi^\prime_1\circ f_1.
\end{equation}
\item[$\bullet$] a skew-symmetric bilinear map $f_2:V_0\times V_0\longrightarrow
V_1^\prime$ satisfying $f_2(\phi_0(x),\phi_0(y))=\phi_1'f_2(x,y)$,
  \end{itemize}
  such that for all $x,y,z\in V_0$ and $m,n\in V_1$, we have
 \begin{itemize}
    \item[$\bullet$] $\dM f_2(x,y)=f_0(l_2(x,y))-l_2'(f_0(x),f_0(y)),$
     \item[$\bullet$] $ f_2(x,\dM m)=f_1(l_2(x,m))-l_2'(f_0(x),f_1(m)),$
      \item[$\bullet$]
      \begin{eqnarray}
\nonumber&&l_2'(f_2(x,y),f_0(\phi_0(z)))+
f_2(l_2(x,y),\phi_0(z))+f_1(l_3(x,y,z))\\
\nonumber&=&l_3^\prime(f_0(x),f_0(y),f_0(z))+l_2'(f_0(\phi_0(x)),f_2(y,z))+l_2'(f_2(x,z),f_0(\phi_0(y)))\\
\label{eqn:morphismphi2}&&+f_2(\phi_0(x),l_2(y,z))+f_2(l_2(x,z),\phi_0(y)).
      \end{eqnarray}
 \end{itemize}
\end{defi}

The identity $HL_\infty$-morphism
$\Id_\huaV:\huaV\longrightarrow\huaV$ has the identity chain map as
it underlying map, together with $(\Id_\huaV)_2=0$, i.e.
$\Id_\huaV=(\Id_{V_0},\Id_{V_1},0)$. Let $\huaV,~\huaV'$ and
$\huaV''$ be $HL_\infty$-algebras, and
 $f : \huaV \rightarrow {\huaV}'$ and ${f}' : {\huaV}' \rightarrow {\huaV}''$ be $HL_{\infty}$-morphisms, we define their composition
 ${f}' \circ f=(({f}'\circ f)_0,({f}'\circ f)_1,({f}'\circ f)_2)$ by
 setting
  $({f}'\circ f)_0 = f'_0 \circ f_0,$
  $({f}'\circ f)_1 = f'_1 \circ f_1,$
   and
   $$
   ({f}'\circ f)_2 (x,y) = f'_2(f_0(x),f_0(y)) +
   f'_1(f_2(x,y)).
   $$
This is exactly the same as the composition of $L_\infty$-morphisms
between 2-term $L_\infty$-algebras. To see that it is indeed a
$HL_\infty$-morphism, we still need to show that the conditions
related with $\phi_0$ and $\phi_1$ in Definition \ref{defi:morphism}
hold. We leave it as an exercise.


Thus, we have
\begin{pro}
  There is a category {\bf 2HL$_\infty$} with 2-term $HL_\infty$-algebras as
  objects and $HL_\infty$-morphisms as morphisms.
\end{pro}

\subsection{The equivalence of hom-Lie 2-algebras and 2-term $HL_\infty$-algebras}

\begin{thm}\label{thm:equivalent}
  The categories {\bf 2HL$_\infty$} and {\bf HLie2} are equivalent.
\end{thm}
\pf We only give a sketch of the proof. First we construct a functor
$T:{\bf 2HL_\infty}\longrightarrow {\bf HLie2}$. Given a 2-term
$HL_\infty$-algebra
$\huaV=(V_1\stackrel{\dM}{\longrightarrow}V_0,l_2,l_3,\phi_0,\phi_1)$,
we have a 2-vector space $L$ given by \eqref{v}. Define the
skew-symmetric bilinear functor $[\cdot,\cdot]:L\times
L\longrightarrow L$ by
$$
[(x,m),(y,n)]=\big(l_2(x,y),l_2(x,n)+l_2(m,y)+l_2(\dM
m,n)\big),\quad\forall~(x,m),(y,n)\in L_1=V_0\oplus V_1.
$$
Define the linear functor $\Phi$ by
$$
\Phi=(\Phi_0,\Phi_1)=(\phi_0,\phi_0\oplus\phi_1).
$$
By the fact that $\phi_0$ and $\phi_1$ commutes with the
differential $\dM$, we deduce that $\Phi$ is a functor, i.e.
$\Phi\in\End(L)$. By Condition (f) and (g) in Definition
\ref{defi:2hl}, we have
\begin{eqnarray*}
  \Phi[(x,m),(y,n)]&=&\big(\phi_0l_2(x,y),\phi_1(l_2(x,n)+l_2(m,y)+l_2(\dM
m,n))\big)\\
&=&\big(l_2(\phi_0(x),\phi_0(y)),l_2(\phi_0(x),\phi_1(n))+l_2(\phi_1(m),\phi_0(y))+l_2(\phi_0(\dM
m),\phi_1(n))\big)\\
&=&\big(l_2(\phi_0(x),\phi_0(y)),l_2(\phi_0(x),\phi_1(n))+l_2(\phi_1(m),\phi_0(y))+l_2(\dM\phi_1(
m),\phi_1(n))\big)\\
&=&[(\phi_0(x),\phi_1(m)),(\phi_0(y),\phi_1(n))]\\
&=&[\Phi(x,m),\Phi(y,n)].
\end{eqnarray*}
Define the Jacobiator by
$$
J_{x,y,z}=([[x,y],\phi_0(z)],l_3(x,y,z)).
$$
It is straightforward to deduce that
\begin{eqnarray*}
  J_{\Phi_0(x),\Phi_0(y),\Phi_0(z)}&=&([[\phi_0(x),\phi_0(y)],\phi_0^2(z)],l_3(\phi_0(x),\phi_0(y),\phi_0(z)))\\
&=&(\phi_0[[x,y],\phi_0(z)],\phi_1l_3(x,y,z))\\
&=&\Phi_1J_{x,y,z}.
\end{eqnarray*}
 By the various conditions of
$(V_1\stackrel{\dM}{\longrightarrow}V_0,l_2,l_3,\phi_0,\phi_1)$
being a 2-term $HL_\infty$-algebra, we deduce that
$(L,[\cdot,\cdot],J,\Phi)$ is a hom-Lie 2-algebra. Thus, we have
constructed a hom-Lie 2-algebra $L=T(\huaV)$ from a 2-term
$HL_\infty$-algebra $\huaV$.

For any $HL_\infty$-morphism $f=(f_0,f_1,f_2)$ form $\huaV$ to
$\huaV'$, next we construct a hom-Lie 2-algebra morphism $F=T(f)$
from $L=T(\huaV)$ to $L'=T(\huaV')$.

Let $F_0=f_0,~F_1=f_0\oplus f_1$, and $F_2$ be given by
$$
F_2(x,y)=([f_0(x),f_0(y)],f_2(x,y)).
$$
Then $F_2(x,y)$ is a bilinear skew-symmetric natural isomorphism
from $[F_0(x),F_0(y)]$ to $F_0[x,y]$, and $F=(F_0,F_1,F_2)$ is a
morphism from $L$ to $L'$.

One can also deduce that $T$ preserves the identity morphisms and
the composition of morphisms. Thus, $T$ constructed above is a
functor from {\bf 2HL$_\infty$} to {\bf HLie2}.

Conversely, given a hom-Lie 2-algebra $L$, we construct the 2-term
$HL_\infty$-algebra $\huaV=S(L)$ as follows. As a complex of vector
spaces, $\huaV$ is obtained by \eqref{eqn:2term}, i.e.
$V_0=L_0,~V_1=\Ker(s)$, and $\dM=t|_{\Ker(s)}$. Define $l_2$ by
$$
l_2(x,y)=[x,y],\quad l_2(x,m)=-l_2(m,x)=[i(x),m],\quad l_2(m,n)=0,
$$
Define $\phi_0=\Phi_0:V_0(=L_0)\longrightarrow V_0$, and define
$\phi_1=\Phi_1|_{V_1=\Ker(s)}:V_1\longrightarrow V_1$. Since $\Phi$
is a functor, we have $\phi_0\circ\dM=\dM\circ\phi_1$. Since $\Phi$
satisfies \eqref{eqn:Phimorphism}, it follows that $\phi_0$ and
$\phi_1$ satisfy Conditions (f) and (g) in Definition
\ref{defi:2hl}.

Furthermore, define $l_3$ by
$$
l_3(x,y,z)=J_{x,y,z}-i(s(J_{x,y,z})).
$$
Since $J_{\Phi_0(x),\Phi_0(y),\Phi_0(z)}=\Phi_1J_{x,y,z}$, we deduce
that $\phi_1l_3(x,y,z)=l_3(\phi_0(x),\phi_0(y),\phi_0(z))$. The
various conditions of $L$ being a hom-Lie 2-algebra imply that
$\huaV$ is 2-term $HL_\infty$-algebra.

Let $F=(F_0,F_1,F_2):L\longrightarrow L'$ be a hom-Lie 2-algebra
morphism, and $S(L)=\huaV,~S(L')=\huaV'$. Define
$S(F)=f=(f_0,f_1,f_2)$ as follows. Let $f_0=F_0$,
$f_1=F_1|_{V_1=\Ker(s)}$ and define $f_2$ by
$$
f_2(x,y)=F_2(x,y)-i(s(F_2(x,y))).
$$
It is not hard to deduce that $f$ is a $HL_\infty$-algebra morphism.
Furthermore, $S$ also preserves the identity morphisms and the
composition of morphisms. Thus, $S$ is a functor from {\bf HLie2} to
{\bf 2HL$_\infty$}.

We are left to show that there are natural isomorphisms
$\alpha:T\circ S\Longrightarrow 1_{{\bf HLie2}}$ and $\beta:S\circ
T\Longrightarrow 1_{{\bf 2HL_\infty}}$. For a hom-Lie 2-algebra
$(L,[\cdot,\cdot],J,\Phi)$, applying the functor $S$ to $L$, we
obtain a 2-term $HL_\infty$-algebra
$\huaV=(V_1=\Ker(s)\stackrel{\dM=t|_{\Ker(s)}}{\longrightarrow}V_0=L_0,l_2,l_3,\phi_0,\phi_1)$.
Applying the functor  $T$ to $\huaV$, we obtain a hom-Lie 2-algebra
$(L',[\cdot,\cdot]',\Phi',J')$, with the space $V_0$ of objects and
the space $V_0\oplus \Ker(s)$ of morphisms. Define
$\alpha_L:L'\longrightarrow L$ by setting
$$
(\alpha_L)_0(x)=x,\quad (\alpha_L)_1(x,m)=i(x)+m.
$$
It is obvious that $\alpha_L$ is an isomorphism of 2-vector spaces.
Furthermore, since $[\cdot,\cdot]$ is a bilinear functor, we have
$[i(x),i(y)]=i([x,y])$, and
$$
[m,n]=[m\cdot_\ve i(\dM m),i(0)\cdot_\ve n]=[m,i(0)]\cdot_\ve[i(\dM
m),n]=[i(\dM m),n].
$$
Therefore, we have
\begin{eqnarray*}
  \alpha_L[(x,m),(y,n)]'&=&\alpha_L(l_2(x,y),l_2(x,n)+l_2(m,y)+l_2(\dM
  m,n))\\
  &=&\alpha_L([x,y],[i(x),n]+[m,i(y)]+[i(\dM m),n])\\
  &=&i([x,y])+[i(x),n]+[m,i(y)]+[i(\dM m),n]\\
  &=&[i(x),i(y)]+[i(x),n]+[m,i(y)]+[m,n]\\
  &=&[i(x)+m,i(y)+n]\\
  &=&[\alpha_L(x,m),\alpha_L(y,n)],
\end{eqnarray*}
which implies that $\alpha_L$ is also a hom-Lie 2-algebra morphism
with $(\alpha_L)_2$ the identity isomorphism. Thus, $\alpha_L$ is an
isomorphism of hom-Lie 2-algebras. It is also easy to see that it is
a natural isomorphism.

For a 2-term $HL_\infty$-algebra
$\huaV=(V_1\stackrel{\dM}{\longrightarrow}V_0,l_2,l_3,\phi_0,\phi_1)$,
applying the functor $S$ to $\huaV$, we obtain a hom-Lie 2-algebra
$(L,[\cdot,\cdot],\Phi)$. Applying the functor $T$ to $L$, we obtain
exactly the same 2-term $HL_\infty$-algebra $\huaV$. Thus,
$\beta_\huaV=\Id_\huaV=(\Id_{V_0},\Id_{V_1})$ is the natural
isomorphism from $T\circ S$ to $1_{{\bf 2HL_\infty}}$. This finishes
the proof. \qed

\begin{rmk}
  We can further obtain 2-categories { $\bf 2HL_\infty$} and  {\bf HLie2}  by
  introducing 2-morphisms and strengthen Theorem
  \ref{thm:equivalent} to the 2-equivalence of 2-categories. Since
  it is a diversion from our aims, we omit the details.
\end{rmk}

\section{Skeletal hom-Lie 2-algebras}

Since we have proved that the category of hom-Lie 2-algebras and the
category of 2-term $HL_\infty$-algebras are equivalent, in the
following, when we say a hom-Lie 2-algebra, what we mean is a 2-term
$HL_\infty$-algebra. In this section, first we give the
classification of hom-Lie 2-algebras, and then we construct examples
of skeletal hom-Lie 2-algebras, which are hom-analogues of string
Lie 2-algebras,  from quadratic hom-Lie algebras introduced in
\cite{BM}.

\subsection{The classification of skeletal hom-Lie 2-algebras}
 A 2-term $HL_\infty$-algebra is called
{\em skeletal} if $\dM=0$. Let $\huaV$ be a skeletal 2-term
$HL_\infty$-algebra. By Condition (h) in Definition \ref{defi:2hl},
we see that $(V_0,l_2(\cdot,\cdot),\phi_0)$ is exactly a hom-Lie
algebra. Define $\rho_{\phi_1}:V_0\longrightarrow\gl(V_1)$ by
\begin{equation}\label{defi:rho}
\rho_{\phi_1}(x)(m)=l_2(x,m),\quad\forall ~x\in V_0,~m\in V_1.
\end{equation}

\begin{pro}\label{pro:representation}
Let $\huaV$ be a skeletal 2-term $HL_\infty$-algebra, then the map
$\rho_{\phi_1}$ defined by \eqref{defi:rho} is a representation of
the hom-Lie algebra $(V_0,l_2(\cdot,\cdot),\phi_0)$ on $V_1$ with
respect to $\phi_1$.
\end{pro}
\pf We only need to check that the two conditions in Definition
\ref{defi:representation} are satisfied. For any $x\in V_0, m\in
V_1$, by Condition (g) in Definition \ref{defi:2hl}, we have
\begin{eqnarray*}
  l_2(\phi_0(x),\phi_1(m))=\phi_1(l_2(x,m)),
\end{eqnarray*}
which means that
$$\rho_{\phi_1}(\phi_0(x))\circ\phi_1=\phi_1\circ\rho_{\phi_1}(x).$$ Thus
Condition (i) in Definition \ref{defi:representation} is satisfied.
Furthermore, since $\huaV$ is skeletal, by Condition (i) we have
$$
l_2(\phi_0(x),l_2(y,m))+l_2(\phi_0(y),l_2(m,x))+l_2(\phi_1(m),l_2(x,y))=0,
$$
which yields that
$$\rho_{\phi_1}(l_2(x,y))\circ\phi_1=\rho_{\phi_1}(\phi_0(x))\circ\rho_{\phi_1}(y)-\rho_{\phi_1}(\phi_0(y))\circ\rho_{\phi_1}(x).$$ Therefore,
Condition (ii) in Definition \ref{defi:representation} is satisfied.
Thus $\rho_{\phi_1}$ is a representation of the hom-Lie algebra
$(V_0,l_2(\cdot,\cdot),\phi_0)$ on $V_1$ with respect to $\phi_1$.
\qed

\begin{thm}
  There is a one-to-one correspondence between skeletal 2-term
  $HL_\infty$-algebras and druples $((\frkg,[\cdot,\cdot]_\frkg,\phi_\frkg),W,A,\rho_A,\theta)$,
  where $(\frkg,[\cdot,\cdot]_\frkg,\phi_\frkg)$ is a hom-Lie algebras, $W$ is a vector space,
  $A\in\gl(W)$, $\rho_A:\frkg\longrightarrow\gl(W)$ is a
  representation of $\frkg$ on $W$ with respect to $A$, and $\theta$
  is a 3-hom-cocycle of the hom-Lie algebra $\frkg$ with
  coefficients in the representation $\rho_A$.
\end{thm}
\pf For any skeletal 2-term $HL_\infty$-algebra $V$,
$(V_0,l_2(\cdot,\cdot),\phi_0)$ is a hom-Lie algebra. By Proposition
\ref{pro:representation}, $\rho_{\phi_1}:V_0\longrightarrow\gl(V_1)$
defined by \eqref{defi:rho} is a representation of the hom-Lie
algebra $(V_0,l_2(\cdot,\cdot),\phi_0)$ on $V_1$ with respect to
$\phi_1$. Now we prove that $l_3$ is a 3-hom-cocycle with respect to
the representation $\rho_{\phi_1}$ and thus any skeletal 2-term
  $HL_\infty$-algebra gives rise to a druple
  $((V_0,l_2(\cdot,\cdot),\phi_0),V_1,\phi_1,\rho_{\phi_1},l_3)$. In fact, by Condition (j) in
  Definition \ref{defi:2hl}, we have
\begin{eqnarray*}
&&l_3(l_2(w,x),\phi_0(y),\phi_0(z))+l_2(l_3(w,x,z),\phi^2_0(y))\\
&&+
l_3(\phi_0(w),l_2(x,z),\phi_0(y))+l_3(l_2(w,z),\phi_0(x),\phi_0(y)) \\
&=&l_2(l_3(w,x,y),\phi^2_0(z))+l_3(l_2(w,y),\phi_0(x),\phi_0(z))+l_3(\phi_0(w),l_2(x,y),\phi_0(z))\\
&&+l_2(\phi^2_0(w),l_3(x,y,z))+l_2(l_3(w,y,z),\phi^2_0(x))+l_3(\phi_0(w),l_2(y,z),\phi_0(x)),
\end{eqnarray*}
which exactly means that
$$
(\dM_{\rho_{\phi_1}} l_3)(w,x,y,z)=0.
$$
 The converse part is easy to be checked and this finishes the
 proof. \qed

 \subsection{The construction of skeletal hom-Lie 2-algebras from quadratic hom-Lie
 algebras}

 \begin{defi}{\rm\cite{BM}}
A quadratic hom-Lie algebra is hom-Lie algebra
$(\frkg,[\cdot,\cdot]_\frkg,\phi_\frkg)$ together with a symmetric
nondegenerate bilinear form
$B:\frkg\times\frkg\longrightarrow\mathbb R$, such that for any
$x,y,z\in\frkg$, the following equalities are satisfied:
\begin{eqnarray}
 \label{eqn:invariant1} B([x,y]_\frkg,z)&=&-B([x,z]_\frkg,y),\\
\label{eqn:invariant2} B(\phi_\frkg(x),y)&=&B(x,\phi_\frkg(y)).
\end{eqnarray}
 \end{defi}

 Recall that a (quadratic) hom-Lie algebra is said to be involutive if
 $\phi_\frkg$ satisfies
 \begin{equation}\label{eqn:involution}
  \phi_\frkg^2=\Id.
 \end{equation}

 For a symmetric nondegenerate bilinear form $B$, there are close
 relations between conditions \eqref{eqn:invariant2},
 \eqref{eqn:involution}, and
 \begin{equation}\label{eqn:invariant3}
   B(\phi_\frkg(x),\phi_\frkg(y))=B(x,y).
 \end{equation}
 \begin{lem}\label{lem:3conditions}
   Let $B$ be a symmetric nondegenerate bilinear form on the hom-Lie
   algebra $(\frkg,[\cdot,\cdot]_\frkg,\phi_\frkg)$. Consider the
   three conditions \eqref{eqn:invariant2}, \eqref{eqn:involution}
   and \eqref{eqn:invariant3}, any two of them can imply the third
   one.
 \end{lem}
\pf If $B$ satisfies \eqref{eqn:invariant2} and
\eqref{eqn:involution}, we have
$$
B(\phi_\frkg(x),\phi_\frkg(y))=B(x,\phi_\frkg^2(y))=B(x,y).
$$

If B satisfies \eqref{eqn:invariant2} and \eqref{eqn:invariant3}, on
one hand, we have $B(\phi_\frkg(x),\phi_\frkg(y))=B(x,y)$. On the
other hand, we have
$B(\phi_\frkg(x),\phi_\frkg(y))=B(x,\phi_\frkg^2(y))$. Thus, we have
$$
B(x,(\phi_\frkg^2-{\Id})(y))=0.
$$
Since $B$ is nondegenerate, we deduce that $\phi_\frkg^2=\Id$.

If  B satisfies \eqref{eqn:involution} and \eqref{eqn:invariant3},
we have
$$
B(\phi_\frkg(x),y)=B(\phi_\frkg(x),\phi_\frkg^2y)=B(x,\phi_\frkg(y)).
$$
This finishes the proof. \qed\vspace{3mm}

 Let $(\frkg,[\cdot,\cdot]_\frkg,\phi_\frkg,B)$ be an involutive
 quadratic hom-Lie algebra. Define $l_3^B:\wedge^3\frkg\longrightarrow\mathbb
 R$ by
\begin{equation}\label{eqn:l3B}
l_3^B(x,y,z)=B([x,y]_\frkg,z).
\end{equation}
By \eqref{eqn:invariant1}, $l_3$ is skew-symmetric.

\begin{lem}\label{lem:l3cocycle}
  $l_3^B$ is a $3$-hom cocycle with coefficients in the trivial
  representation.
\end{lem}
\pf First, by Lemma \ref{lem:3conditions}, we have
\begin{eqnarray*}
l_3^B(\phi_\frkg(x),\phi_\frkg(y),\phi_\frkg(z))&=&B([\phi_\frkg(x),\phi_\frkg(y)]_\frkg,\phi_\frkg(z))=B(\phi_\frkg[x,y]_\frkg,\phi_\frkg(z))\\
&=&B([x,y]_\frkg,z)=l_3^B(x,y,z),
\end{eqnarray*}
which implies that $l_3$ is a $3$-hom-cochain. Furthermore, by
\eqref{eqn:invariant1} and the hom-Jacobi identity, we have
\begin{eqnarray*}
 && 2\dM_Tl_3^B(w,x,y,z)\\
 &=&2\big(-l_3^B([w,x]_\frkg,\phi_\frkg(y),\phi_\frkg(z))+l_3^B([w,y]_\frkg,\phi_\frkg(x),\phi_\frkg(z))-l_3^B([w,z]_\frkg,\phi_\frkg(x),\phi_\frkg(y))\\
 &&-l_3^B([x,y]_\frkg,\phi_\frkg(w),\phi_\frkg(z))+l_3^B([x,z]_\frkg,\phi_\frkg(w),\phi_\frkg(y))-l_3^B([y,z]_\frkg,\phi_\frkg(w),\phi_\frkg(x))\big)\\
 &=&-B([[w,x]_\frkg,\phi_\frkg(y)]_\frkg,\phi_\frkg(z))+B([[w,y]_\frkg,\phi_\frkg(x)]_\frkg,\phi_\frkg(z))-B([[w,z]_\frkg,\phi_\frkg(x)]_\frkg,\phi_\frkg(y))\\
 &&-B([[x,y]_\frkg,\phi_\frkg(w)]_\frkg,\phi_\frkg(z))+B([[x,z]_\frkg,\phi_\frkg(w)]_\frkg,\phi_\frkg(y))-B([[y,z]_\frkg,\phi_\frkg(w)]_\frkg,\phi_\frkg(x))\\
 &&+B([[w,x]_\frkg,\phi_\frkg(z)]_\frkg,\phi_\frkg(y))-B([[w,y]_\frkg,\phi_\frkg(z)]_\frkg,\phi_\frkg(x))+B([[w,z]_\frkg,\phi_\frkg(y)]_\frkg,\phi_\frkg(x))\\
 &&+B([[x,y]_\frkg,\phi_\frkg(z)]_\frkg,\phi_\frkg(w))-B([[x,z]_\frkg,\phi_\frkg(y)]_\frkg,\phi_\frkg(w))+B([[y,z]_\frkg,\phi_\frkg(x)]_\frkg,\phi_\frkg(w))\\
 &=&0.
\end{eqnarray*}
Thus, $l_3^B$ is a $3$-hom-cocycle.\qed\vspace{3mm}

Now we are ready to construct an example of skeletal hom-Lie
2-algebras
$\huaV=(V_1\stackrel{0}{\longrightarrow}V_0,l_2,l_3,\phi_0,\phi_1)$
from an involutive quadratic hom-Lie algebra
$(\frkg,[\cdot,\cdot]_\frkg,\phi_\frkg,B)$ as follows. Let
$V_1=\mathbb R$, $V_0=\frkg$, $\phi_0=\phi_\frkg$ and $\phi_1=\Id$.
Define $l_2$  by
\begin{equation}\label{eqn:l_2string}
l_2(x,y)=[x,y]_\frkg,\quad l_2(x,m)=0,
\end{equation}
and define $l_3$ by \eqref{eqn:l3B}. By Lemma \ref{lem:l3cocycle},
it is straightforward to see that all the conditions in Definition
\ref{defi:2hl} are satisfied. Therefore $(\mathbb
R\stackrel{0}{\longrightarrow}\frkg,l_2,l_3^B,\phi_\frkg,\Id)$ is a
skeletal hom-Lie 2-algebra for any  involutive quadratic hom-Lie
algebra $(\frkg,[\cdot,\cdot]_\frkg,\phi_\frkg,B)$.

In the following, we construct the hom-analogue of string Lie
2-algebras. We need some preparations. For any involutive hom-Lie
algebra $(\frkg,[\cdot,\cdot]_\frkg,\phi_\frkg)$,
$(\frkg_{\phi_\frkg},[\cdot,\cdot]_{\phi_\frkg})$ is a Lie algebra
\cite[Theorem 5.1]{BM}, where $[\cdot,\cdot]_{\phi_\frkg}$ is given
by
$$
[x,y]_{\phi_\frkg}=[\phi_\frkg(x),\phi_\frkg(y)]_\frkg=\phi_\frkg([x,y]_\frkg).
$$
\begin{thm}\label{thm:cohomologyrelation}
  There is an inclusion from $H^k(\frkg)$ to
  $H^k(\frkg_{\phi_\frkg})$, where $H^k(\frkg)$ is the $k$-th
  cohomology group of the hom-Lie algebra
  $(\frkg,[\cdot,\cdot]_\frkg,\phi_\frkg)$ with the coefficients in
  the trivial representation, and
  $H^k(\frkg_{\phi_\frkg})$ is the $k$-th
  cohomology group of the Lie algebra
  $(\frkg_{\phi_\frkg},[\cdot,\cdot]_{\phi_\frkg})$ with the coefficients
  in the trivial representation.
\end{thm}
\pf We only need to show that for any $f\in
Z^k_{\phi_\frkg}(\frkg)$, as a $k$-cochain of $\frkg_{\phi_\frkg}$,
$f$ is also closed, and for any $f\in B^k_{\phi_\frkg}(\frkg)$, as a
$k$-cochain of $\frkg_{\phi_\frkg}$, $f$ is also exact. In fact, for
any $f\in Z^k_{\phi_\frkg}(\frkg)$, we have
\begin{eqnarray*}
  f(\phi_\frkg(x_1),\cdots,\phi_\frkg(x_k))&=&f(x_1,\cdots,x_k),\\
  \dM_Tf(x_1,\cdots,x_{k+1})&=&\sum_{i<j}(-1)^{i+j}f([x_i,x_j]_\frkg,\phi_\frkg(x_1),\cdots,\widehat{x_i},\cdots,\widehat{x_j},\cdots,\phi_\frkg(x_{k+1}))=0.
\end{eqnarray*}
Since $\phi_\frkg^2=\Id$, we have
\begin{eqnarray*}
  0&=&\sum_{i<j}(-1)^{i+j}f([\phi_\frkg^2(x_i),\phi_\frkg^2(x_j)]_\frkg,\phi_\frkg(x_1),\cdots,\widehat{x_i},\cdots,\widehat{x_j},\cdots,\phi_\frkg(x_{k+1}))\\
  &=&\sum_{i<j}(-1)^{i+j}f(\phi_\frkg[\phi_\frkg(x_i),\phi_\frkg(x_j)]_\frkg,\phi_\frkg(x_1),\cdots,\widehat{x_i},\cdots,\widehat{x_j},\cdots,\phi_\frkg(x_{k+1}))\\
  &=&\sum_{i<j}(-1)^{i+j}f([\phi_\frkg(x_i),\phi_\frkg(x_j)]_\frkg,x_1,\cdots,\widehat{x_i},\cdots,\widehat{x_j},\cdots,x_{k+1})\\
  &=&\sum_{i<j}(-1)^{i+j}f([x_i,x_j]_{\phi_\frkg},x_1,\cdots,\widehat{x_i},\cdots,\widehat{x_j},\cdots,x_{k+1})\\
  &=&\dM_{\frkg_{\phi_\frkg}}f(x_1,\cdots,x_{k+1}),
\end{eqnarray*}
where $\dM_{\frkg_{\phi_\frkg}}$ is the coboundary operator of the
Lie algebra $\frkg_{\phi_\frkg}$ with the coefficients in the
trivial representation. Therefore, as a $k$-cochain of
$\frkg_{\phi_\frkg}$, $f$ is also closed.

For any $f\in B^k_{\phi_\frkg}(\frkg)$, assume that $f=\dM_Th$, for
some $h:\wedge^{k-1}\frkg\longrightarrow\mathbb R$ satisfying
$h\circ\phi_\frkg=h$. Similar as the above proof, we have
\begin{eqnarray*}
  f(x_1,\cdots,x_{k})=\dM_Th(x_1,\cdots,x_k)
=\dM_{\frkg_{\phi_\frkg}}h(x_1,\cdots,x_{k}),
\end{eqnarray*}
which implies that, as a $k$-cochain of $\frkg_{\phi_\frkg}$, $f$ is
also exact. This finishes the proof. \qed\vspace{2mm}

Now let the involutive hom-Lie algebra
$(\frkg,[\cdot,\cdot]_\frkg,\phi_\frkg)$ be semisimple\footnote{For
the notion of semisimple hom-Lie algebras, we refer to \cite{BM} and
references therein}, then the Lie algebra
$(\frkg_{\phi_\frkg},[\cdot,\cdot]_{\phi_\frkg})$ is also semisimple
\cite{BM}. Furthermore, the authors define a symmetric bilinear form
$B:\frkg\times \frkg\longrightarrow\mathbb R$ by
\begin{equation}\label{eqn:BKilling}
B(x,y)=\tr(\ad_x\circ\ad_y),
\end{equation}
where $\ad_x$ is defined as usual: $\ad_xy=[x,y]_\frkg$, and
$(\frkg,[\cdot,\cdot]_\frkg,\phi_\frkg, B)$ is a semisimple
quadratic involutive hom-Lie algebra. There is also the following
relation
$$
K_{\frkg_{\phi_\frkg}}(x,y)=B(\phi_\frkg(x),y),
$$
where $K_{\frkg_{\phi_\frkg}}$ is the Killing form of the semisimple
Lie algebra $\frkg_{\phi_\frkg}$.

\begin{cor}
  Let $(\frkg,[\cdot,\cdot]_\frkg,\phi_\frkg)$ be a semisimple involutive hom-Lie
  algebra, then the cohomology class of $l_3^B$ defined by
  \eqref{eqn:l3B} is not trivial, where $B$ is given by \eqref{eqn:BKilling}.
\end{cor}
\pf Since $\phi_\frkg^2=\Id$, we have
 \begin{eqnarray*}
    K([x,y]_{\phi_\frkg},z)=B(\phi_\frkg[x,y]_{\phi_\frkg},z)=B([\phi_\frkg^2(x),\phi_\frkg^2(y)]_\frkg,z)=B([x,y]_\frkg,z)=l_3^B(x,y,z).
  \end{eqnarray*}
Thus, $l_3^B$ defined by
  \eqref{eqn:l3B}, as a $3$-cochain of the Lie algebra $\frkg_{\phi_\frkg}$, is exactly the Cartan $3$-form of
  $\frkg_{\phi_\frkg}$. By Theorem
  \ref{thm:cohomologyrelation}, if $l_3^B$ is exact, we deduce that
  the Cartan $3$-form of the semisimple Lie  algebra
  $\frkg_{\phi_\frkg}$ is exact, this is a conflict.\qed

  \begin{defi}
  The hom-analogue of the string Lie 2-algebra associated to any semisimple involutive hom-Lie
  algebra $(\frkg,[\cdot,\cdot]_\frkg,\phi_\frkg)$ is the hom-Lie 2-algebra  $(\mathbb
R\stackrel{0}{\longrightarrow}\frkg,l_2,l_3^B,\phi_\frkg,\Id)$,
where $l_2,~l_3^B$ and $B$ are given by \eqref{eqn:l_2string},
\eqref{eqn:l3B} and \eqref{eqn:BKilling} respectively.
  \end{defi}

\begin{ex}
  Consider the semisimple Lie algebra  $\sln(2)$, with basis $A=\Big(\begin{array}{cc}0&1\\
  0&0\end{array}\Big)$, $B=\Big(\begin{array}{cc}0&0\\
  1&0\end{array}\Big)$, and $C=\Big(\begin{array}{cc}1&0\\
  0&-1\end{array}\Big)$ satisfying the relation
  $$
[A,B]=C,\quad [C,A]=2A,\quad [B,C]=2B.
  $$
  For any $x\in\sln(2)$, let $\phi(x)=-x^T$, the minus of the
  transpose of $x$. Obviously, $\phi$ is an involution map. Then
  $(\sln(2),[\cdot,\cdot]_\phi,\phi)$ is a semisimple involutive hom-Lie
  algebra. More precisely, we have
  \begin{equation}\label{eqn:exl2}
[A,B]_\phi=[\phi(A),\phi(B)]=[-B,-A]=-C,\quad [C,A]_\phi=-2B,\quad
[B,C]_\phi=-2A.
\end{equation}
It is easy to obtain that
 \begin{equation}\label{eqn:exl3}
l_3^B(A,B,C)=B([A,B]_\phi,C)=-\tr(\ad_C^2)=8.
\end{equation}
Therefore, we obtain a hom-analogue of the string Lie 2-algebra
$(\mathbb
R\stackrel{0}{\longrightarrow}\sln(2),l_2,l_3^B,\phi,\Id)$, where
$l_2$ and $l_3^B$ are determined by \eqref{eqn:exl2} and
\eqref{eqn:exl3}.
\end{ex}

\section{Strict hom-Lie 2-algebras}

In this section, we introduce the notion of crossed modules of
hom-Lie algebras, and we prove that there is a one-to-one
correspondence between crossed modules of hom-Lie algebras and
strict hom-Lie 2-algebras. Here what we mean a strict hom-Lie
2-algebra is a 2-term $HL_\infty$-algebra whose $l_3$ is zero. Then
we construct strict hom-Lie 2-algebras from hom-left-symmetric
algebras. At last, we introduce the notion of symplectic hom-Lie
algebras, and give the construction of strict hom-Lie 2-algebras
from symplectic hom-Lie algebras.

\subsection{Strict hom-Lie 2-algebras and crossed modules of hom-Lie algebras}
  \begin{defi}
A crossed module of hom-Lie algebras is a quadruple
$((\frkh,\br__\frkh,\phi_\frkh),(\frkg,\br__\frkg,\phi_\frkg),dt,\varphi)$,
where $(\frkh,\br__\frkh,\phi_\frkh)$ and
$(\frkg,\br__\frkg,\phi_\frkg)$ are hom-Lie algebras,
$dt:\frkh\longrightarrow\frkg$ is a hom-Lie algebra morphism and
$\varphi$ is a representation of the hom-Lie algebra $\frkg$ on
$\frkh$, such that
\begin{eqnarray}\label{eqn:con1cm}
dt(\varphi_x(m))&=&[x,dt(m)]_{\frkg},\\\label{eqn:con2cm}
\varphi_{dt(m)}(m^\prime)&=&[m,m^\prime]_{\frkh}.
\end{eqnarray}
\end{defi}

\begin{lem}
  Let $((\frkh,\phi_\frkh),(\frkg,\phi_\frkg),dt,\varphi)$ be a  crossed module of
  hom-Lie  algebras, then we have
  \begin{equation}
    \varphi_{\phi_\frkg(x)}([m,n]_\frkh)=[\varphi_xm,\phi_\frkh(n)]_\frkh+[\phi_\frkh
    (m),\varphi_xn]_\frkh.
  \end{equation}
\end{lem}
\pf By the fact that $\varphi$ is a representation, we have
$$
\varphi_{[x,y]_\frkg}\circ\phi_\frkh=\varphi_{\phi_\frkg
(x)}\circ\varphi_y-\varphi_{\phi_\frkg (y)}\circ\varphi_x.
$$
Let $y=dt(m)$, by \eqref{eqn:con1cm} and \eqref{eqn:con2cm}, we
obtain
$$
\varphi_{dt(\varphi_xm)}\phi_\frkh (n)=\varphi_{\phi_\frkg
x}\circ\varphi_{dt(m)}n-\varphi_{\phi_\frkg(dt(m))}\circ\varphi_xn,
$$
which implies that
$$
[\varphi_xm,\phi_\frkh (n)]_\frkh=\varphi_{\phi_\frkg
(x)}[m,n]_\frkh-[\phi_\frkh (m),\varphi_xn]_\frkh. \qed
$$

\begin{rmk}
  If $\phi_\frkg=\Id$ and $\phi_\frkh=\Id$, i.e.
  $(\frkh,\frkg,dt,\varphi)$ is a crossed module of Lie algebras, we
  deduce that $\varphi$ must act as a derivation by the above proof.
\end{rmk}

\begin{thm}
  There is a one-to-one correspondence between strict hom-Lie
  2-algebras and crossed modules of hom-Lie algebras.
\end{thm}

\pf Let
$(V_1\stackrel{\dM}{\longrightarrow}V_0,l_2,l_3=0,\phi_0,\phi_1)$ be
a strict hom-Lie 2-algebra, we construct a crossed module of hom-Lie
algebra as follows. Let $\frkg=V_0$ with the bracket operation
$[\cdot,\cdot]_\frkg=l_2:V_0\times V_0\longrightarrow V_0,$ and
linear transformation $\phi_\frkg=\phi_0$. Let $\frkh=V_1$ with the
bracket operation $[\cdot,\cdot]_\frkh:V_1\times V_1\longrightarrow
V_1$ given by
$$
[m,n]_\frkh=l_2(\dM m,n),
$$
and linear transformation $\phi_\frkh=\phi_1$. Furthermore, let
$dt=\dM$.

By (a), (f) and (h), it is obvious that
$(\frkg,[\cdot,\cdot]_\frkg,\phi_0)$ is a hom-Lie algebra. By (b)
and (e), the bracket operation $[\cdot,\cdot]_\frkh$ is well
defined. By (g), we have
$$
\phi_1([m,n]_\frkh)=\phi_1(l_2(\dM m,n))=l_2(\phi_0(\dM
m),\phi_1(n))=l_2(\dM\circ\phi_1 (m),\phi_1(n))=[\phi_1
(m),\phi_1(n)]_\frkh,
$$
which implies that $\phi_1$ is an algebra morphism with respect to
$[\cdot,\cdot]_\frkh$. By (i), we have
\begin{eqnarray*}
 && [\phi_1(m),[n,p]_\frkh]_\frkh+[\phi_1(n),[p,m]_\frkh]_\frkh+[\phi_1(p),[m,n]_\frkh]_\frkh\\
 &=&l_2(\dM\circ\phi_1(m),l_2(\dM n,p))+ l_2(\dM\circ\phi_1(n),l_2(\dM p,m))+ l_2(\dM\circ\phi_1(p),l_2(\dM
 m,n))\\
 &=&l_2(\phi_0(\dM m),l_2(\dM n,p))+ l_2(\phi_0( \dM n),l_2(\dM p,m))+ l_2(\phi_0(\dM p),l_2(\dM
 m,n))\\
 &=&l_2(\phi_0(\dM m),l_2(\dM n,p))+ l_2(\phi_0( \dM n),l_2( p,\dM m))+ l_2(\phi_1( p),l_2(\dM
 m,\dM n))\\
&=&0.
\end{eqnarray*}
Thus, $(\frkh,[\cdot,\cdot]_\frkh,\phi_1)$ is a hom-Lie algebra. By
(d), it is obvious that $dt$ is a morphism of hom-Lie algebras. At
last, define $\varphi:\frkg\times\frkh\longrightarrow\frkh$ by
$$
\varphi_xm=l_2(x,m).
$$
By (g), we have
$\varphi_{\phi_\frkg(x)}\phi_\frkh(m)=\phi_\frkh\varphi_xm$. By (i),
we have
\begin{eqnarray*}
&&\varphi_{[x,y]_\frkg}\phi_\frkh(m)-\varphi_{\phi_\frkg(x)}\circ\varphi_{y}m+\varphi_{\phi_\frkg(y)}\circ\varphi_{x}m\\
&=&l_2(l_2(x,y),\phi_1(m))-l_2(\phi_0(x),l_2(y,m))+l_2(\phi_0(y),l_2(x,m))=0.
\end{eqnarray*}
Thus, $\varphi$ is a representation.  By (d), we see that the
equality \eqref{eqn:con1cm} holds. By the definition of $\varphi$
and $[\cdot,\cdot]_\frkh$, it is obvious that the equality
\eqref{eqn:con2cm} holds. Therefore,
$((\frkh,[\cdot,\cdot]_\frkh,\phi_\frkh),(\frkg,[\cdot,\cdot]_\frkg,\phi_\frkg),dt,\varphi)$
is a crossed module of hom-Lie algebras.

Conversely, given a crossed module of hom-Lie algebras
$((\frkh,[\cdot,\cdot]_\frkh,\phi_\frkh),(\frkg,[\cdot,\cdot]_\frkg,\phi_\frkg),dt,\varphi)$,
we obtain a strict hom-Lie 2-algebra as follows. Let
$V_0=\frkg,~\phi_0=\phi_\frkg,~V_1=\frkh,~\phi_1=\phi_\frkh$ and
$\dM=dt$. Define $l_2:V_i\times V_j\longrightarrow V_{i+j}$ by
$$
l_2(x,y)=[x,y]_\frkg,\quad l_2(x,m)=-l_2(m,x)=\varphi_xm,\quad
l_2(m,n)=0.
$$
The crossed module structure gives various conditions of strict
hom-Lie 2-algebras. We omit the details. \qed\vspace{3mm}

First we have the following trivial example of strict Lie
2-algebras.

\begin{ex}
For any hom-Lie algebra $(\frkg,[\cdot,\cdot]_\frkg,\phi_\frkg)$,
$(\frkg\stackrel{0}{\longrightarrow}\frkg,l_2=[\cdot,\cdot]_\frkg,\phi_0=\phi_\frkg,\phi_1=\phi_\frkg)$
is a strict hom-Lie 2-algebra.
\end{ex}

\subsection{The construction of strict hom-Lie 2-algebras from hom-left-symmetric algebras}

Hom-left-symmetric algebras, or hom-pre-Lie algebras were first
introduced in \cite{MS2}, and then further studied in \cite{Yao3}
and \cite{BaiHom}.

\begin{defi}
  A hom-left-symmetric algebra is a triple $(V,\star,\phi)$, where
  $V$ is a vector space, $\star:V\times V\longrightarrow V$ is a
  bilinear map, and $\phi\in\gl(V)$ such that the following
  equalities are satisfied:
  \begin{eqnarray}
    \label{eqn:conleft1}\phi( x\star y)&=&\phi (x)\star \phi(y),\\
    \label{eqn:conleft2}\phi(x)\star(y\star z)-(x\star y)\star \phi(z)&=&\phi(y)\star(x\star z)-(y\star x)\star
    \phi(z).
  \end{eqnarray}
\end{defi}

Let $(V,\star,\phi)$ be a hom-left-symmetric algebra, define
$[\cdot,\cdot]_V:V\wedge V\longrightarrow V$ by
\begin{equation}
[x,y]_V=x\star y-y\star x,
\end{equation}
and define $\rho_{\phi}:V\longrightarrow\gl(V)$ by
\begin{equation}\label{eqn:repleft}
\rho_\phi(x)(y)=x\star y.
\end{equation}

\begin{pro}\label{pro:homLiefromhomleft}
  With the above notations, $(V,[\cdot,\cdot]_V,\phi)$ is a hom-Lie
  algebra, which is called the sub-adjacent hom-Lie algebra of the hom-left-symmetric algebra $(V,\star,\phi)$, and $\rho_\phi$ is a representation of the hom-Lie
  algebra $V$ on the vector space $V$ with respect to $\phi$. Moreover, if $\phi^2=\Id$, i.e. $(V,[\cdot,\cdot]_V,\phi)$ is an
involutive hom-Lie
  algebra, then the map $\rho^*_\phi:V\longrightarrow\gl(V^*)$
  defined by
  $$
\langle\rho^*_\phi(x)(\xi),y\rangle=-\langle\xi,\rho_\phi(x)(y)\rangle,
  $$
  is also a representation of $(V,[\cdot,\cdot]_V,\phi)$ on the
  vector space $V^*$ with respect to $\phi^*$.
\end{pro}

\pf The first part follows from straightforward computations. As for
the second part, first we should note that, in general, for a
representation $\rho_A$ of the hom-Lie algebra
$(\frkg,[\cdot,\cdot]_\frkg,\phi_\frkg)$ on the vector space $U$
with respect to $A\in\gl(U)$, the induced map
$\rho_A^*:\frkg\longrightarrow\gl(U^*)$,
$\langle\rho_A^*(x)\xi,u\rangle=-\langle\xi,\rho_A(x)(u)\rangle$,
for any $\xi\in U^*$ and $u\in U$, is a representation iff
(\cite[Proposition 2.5]{BM})
$$
A\circ
\rho_A([x,y]_\frkg)=\rho_A(x)\circ\rho_A(\phi_\frkg(y))-\rho_A(y)\circ\rho_A(\phi_\frkg(x)).
$$
Consider the sub-adjacent hom-Lie algebra
$(V,[\cdot,\cdot]_V,\phi)$, if $\phi^2=\Id$, the above condition
reduces to
\begin{eqnarray*}
(\phi(x)\star\phi(y))\star\phi(z)-(\phi(y)\star\phi(x))\star\phi(z)&=&x\star(\phi(y)\star
z)-y\star(\phi(x)\star z)\\
&=&\phi^2(x)\star(\phi(y)\star z)-\phi^2(y)\star(\phi(x)\star z),
\end{eqnarray*}
which holds naturally by \eqref{eqn:conleft2}. This finishes the
proof.\qed\vspace{2mm}

The following procedure provides a way to construct examples of
strict hom-Lie 2-algebras from hom-left-symmetric algebras.

\begin{pro}\label{pro:procedure1}
  Let $(V,\star,\phi)$ be a hom-left-symmetric algebra, for any linear map $\dM:V\longrightarrow V$ satisfying
  \begin{eqnarray}
 \label{eqn:dhom1} \dM\circ\phi&=&\phi\circ\dM,\\
 \label{eqn:dhom2} \dM x\star y&=&x\star\dM y,\\
   \label{eqn:dhom3} \dM(x\star y)&=&x\star\dM y-\dM y\star x,
  \end{eqnarray}
  define $l_2$ on the 2-term complex of vector spaces
  $V_1=V\stackrel{\dM}{\longrightarrow}V_0=V$ by
\begin{equation}\label{eqn:homl2}
\left\{\begin{array}{rlll}l_2(x,y)&=&[x,y]_V, &\forall ~x,y\in V_0\\
l_2(x,y)&=&-l_2(y,x)=x\star y, &\forall ~x\in V_0, ~y\in V_1\\
l_2(x,y)&=&0, &\forall~ x,y\in V_1.\end{array}\right.
\end{equation}
Then
$(V_1=V\stackrel{\dM}{\longrightarrow}V_0=V,l_2,\phi_0=\phi,\phi_1=\phi)$
is a strict hom-Lie 2-algebra.
\end{pro}
\pf By the definition of  $l_2$, conditions (a), (b) and (c) in
Definition \ref{defi:2hl} are satisfied obviously. By Proposition
\ref{pro:homLiefromhomleft}, conditions (f), (g), (h) and (i) are
also satisfied. At last, \eqref{eqn:dhom2} and \eqref{eqn:dhom3}
imply that conditions (d) and (e) hold. \qed


\subsection{The construction of strict hom-Lie 2-algebras from  symplectic hom-Lie algebras}

\begin{defi}
 Let $(\frkg,[\cdot,\cdot]_\frkg,\phi_\frkg)$ be a regular hom-Lie algebra,
 $\omega\in\wedge^2\frkg^*$ is called a symplectic structure on $\frkg$ if
 \begin{itemize}
\item $\omega$ is nondegenerate, i.e. the induced skewsymmetric map
$\omega^\sharp:\frkg\longrightarrow
 \frkg^*$,
$ \langle\omega^\sharp(x),y\rangle=\omega(x,y), $ is nondegenerate;
\item $\omega$ is a 2-hom-cocycle, i.e. we have
$\omega\circ\phi_\frkg=\omega$, and $\dM_T\omega=0$:
\begin{equation}
  \omega(\phi_\frkg(x),[y,z]_\frkg)+\omega(\phi_\frkg(y),[z,x]_\frkg)+\omega(\phi_\frkg(z),[x,y]_\frkg)=0.
\end{equation}
\end{itemize}
$(\frkg,\omega)$ is called a symplectic hom-Lie algebra if $\omega $
is a symplectic structure on $\frkg$.
\end{defi}

Define a bilinear map $\star:\frkg\times\frkg\longrightarrow\frkg$
on the regular symplectic hom-Lie algebra $(\frkg,\omega)$  by
\begin{equation}\label{eqn:defistar}
\omega(x\star y,\phi_\frkg(z))=-\omega(\phi_\frkg(y),[x,z]_\frkg).
\end{equation}
By the fact $\omega$ is closed, we have
$$
\omega(x\star y-y\star
x,\phi_\frkg(z))=-\omega(\phi_\frkg(y),[x,z]_\frkg)+\omega(\phi_\frkg(x),[y,z]_\frkg)=\omega([x,y]_\frkg,\phi_\frkg(z)),
$$
which implies that
\begin{equation}\label{eqn:subadjacent}
 [x,y]_\frkg=x\star y-y\star x,
\end{equation}
since $\phi_\frkg$ is nondegenerate.

\begin{pro}\label{pro:inducedleft}
  Let $((\frkg,[\cdot,\cdot]_\frkg,\phi_\frkg),\omega)$ be a regular symplectic hom-Lie algebra, then
  $(\frkg,\star, \phi_\frkg)$ is a hom-left-symmetric algebra.
  Furthermore, the hom-Lie algebra $\frkg$ is its sub-adjacent hom-Lie
  algebra.
\end{pro}
\pf By \eqref{eqn:defistar}, we have
\begin{eqnarray*}
\omega(\phi_\frkg(x)\star
\phi_\frkg(y),\phi_\frkg^2(z))&=&-\omega(\phi_\frkg^2(y),[\phi_\frkg(x),\phi_\frkg(z)]_\frkg)=-\omega(\phi_\frkg(y),[x,z]_\frkg)\\
&=&\omega(x\star y,\phi_\frkg(z))=\omega(\phi_\frkg(x\star
y),\phi_\frkg^2(z)).
\end{eqnarray*}
Since both $\omega$ and $\phi_\frkg$ are nondegenerate, we deduce
that $\phi_\frkg(x\star y)=\phi_\frkg(x)\star \phi_\frkg(y)$. We
have
\begin{eqnarray*}
  &&\omega\big(\phi_\frkg(x)\star(y\star z)-(x\star y)\star \phi_\frkg(z)-\phi_\frkg(y)\star(x\star z)+(y\star x)\star
  \phi_\frkg(z),\phi^2_\frkg(w)\big)\\
    &=&-\omega(\phi_\frkg(y\star
    z),[\phi_\frkg(x),\phi_\frkg(w)]_\frkg)+\omega(\phi_\frkg^2(z),[x\star
    y,\phi_\frkg(w)]_\frkg)\\
    &&+\omega(\phi_\frkg(x\star
    z),[\phi_\frkg(y),\phi_\frkg(w)]_\frkg)-\omega(\phi_\frkg^2(z),[y\star
    x,\phi_\frkg(w)]_\frkg)\\
&=&-\omega(\phi_\frkg(y)\star
    \phi_\frkg(z),\phi_\frkg[x,w]_\frkg)+\omega(\phi_\frkg^2(z),[[x,
    y]_\frkg,\phi_\frkg(w)]_\frkg)+\omega(\phi_\frkg(x)\star
    \phi_\frkg(z),\phi_\frkg[y,w]_\frkg)\\
    &=&\omega(
    \phi_\frkg^2(z),[\phi_\frkg(y),[x,w]_\frkg]_\frkg)+\omega(\phi_\frkg^2(z),[[x,
    y]_\frkg,\phi_\frkg(w)]_\frkg)-\omega(
    \phi_\frkg^2(z),[\phi_\frkg(x),[y,w]_\frkg]_\frkg)\\
    &=&0,
\end{eqnarray*}
which implies that
$$
\phi_\frkg(x)\star(y\star z)-(x\star y)\star
\phi_\frkg(z)-\phi_\frkg(y)\star(x\star z)+(y\star x)\star
  \phi_\frkg(z)=0.
$$
Thus, $(\frkg,\star, \phi_\frkg)$ is a hom-left-symmetric algebra.
By \eqref{eqn:subadjacent}, the second conclusion is obvious.
\qed\vspace{2mm}

The following theorem provides a procedure to construct strict
hom-Lie 2-algebras from involutive symplectic hom-Lie algebras.

\begin{thm}
 Let $((\frkg,[\cdot,\cdot]_\frkg,\phi_\frkg),\omega)$ be an involutive symplectic hom-Lie algebra, and
  $(\frkg,\star, \phi_\frkg)$ be the induced hom-left-symmetric
  algebra as in Proposition \ref{pro:inducedleft}. On the complex of
  vector spaces
  $\frkg^*\stackrel{\phi_\frkg\circ(\omega^\sharp)^{-1}}{\longrightarrow}\frkg$,
  define $l_2$ by
  \begin{equation}\label{eqn:homl2final}
\left\{\begin{array}{rlll}l_2(x,y)&=&[x,y]_\frkg, &\forall ~x,y\in \frkg\\
l_2(x,\xi)&=&-l_2(\xi,x)=\rho_{\phi_\frkg}^* (x)\xi, &\forall ~x\in \frkg, ~\xi\in \frkg^*\\
l_2(\xi,\eta)&=&0, &\forall~ \xi,\eta\in \frkg^*,\end{array}\right.
\end{equation}
where $\rho_{\phi_\frkg}^* $ is the dual representation of
$\rho_{\phi_\frkg}$ given by \eqref{eqn:repleft}, then
$(\frkg^*\stackrel{\phi_\frkg\circ(\omega^\sharp)^{-1}}{\longrightarrow}\frkg,l_2,\phi_0=\phi_\frkg,\phi_1=\phi_\frkg^*)$
is a strict hom-Lie 2-algebra.
\end{thm}
\pf Similar as the proof of Proposition \ref{pro:procedure1}, we
only need to show that
\begin{eqnarray}
\label{eqn:confinal1}\phi_\frkg\circ(\omega^\sharp)^{-1}\circ\phi_\frkg^*&=&\phi_\frkg^2\circ(\omega^\sharp)^{-1},\\
\label{eqn:confinal2}\phi_\frkg\circ(\omega^\sharp)^{-1}l_2(x,\xi)&=&l_2(x,\phi_\frkg\circ(\omega^\sharp)^{-1}(\xi)),\\
\label{eqn:confinal3}l_2(\phi_\frkg\circ(\omega^\sharp)^{-1}(\xi),\eta)&=&l_2(\xi,\phi_\frkg\circ(\omega^\sharp)^{-1}(\eta)).
\end{eqnarray}
The equality \eqref{eqn:confinal1} is equivalent to
$$
\langle(\omega^\sharp)^{-1}\circ\phi_\frkg^*\xi,\eta\rangle=\langle\phi_\frkg\circ(\omega^\sharp)^{-1}\xi,\eta\rangle.
$$
Let $\xi=\omega^\sharp(x)$ and $\eta=\omega^\sharp(y)$, since
$\omega$ is skew-symmetric, we have
\begin{eqnarray*}
\langle(\omega^\sharp)^{-1}\circ\phi_\frkg^*\xi,\eta\rangle&=&-\langle\phi_\frkg^*(\omega^\sharp(x)),y\rangle
=-\langle\omega^\sharp(x),\phi_\frkg(y)\rangle=\omega(\phi_\frkg(y),x),\\
\langle\phi_\frkg\circ(\omega^\sharp)^{-1}\xi,\eta\rangle&=&\langle\phi_\frkg(x),\omega^\sharp(y)\rangle=\omega(y,\phi_\frkg(x)).
\end{eqnarray*}
Since $\omega\circ\phi_\frkg=\omega$ and $\phi_\frkg^2=\Id$, by
Lemma \ref{lem:3conditions}, we deduce that
$\omega(y,\phi_\frkg(x))=\omega(\phi_\frkg(y),x)$. Therefore,
\eqref{eqn:confinal1} holds.

The equality \eqref{eqn:confinal2} is equivalent to
$$
\langle(\omega^\sharp)^{-1}\circ\phi_\frkg^*l_2(x,\xi),\eta\rangle=\langle
l_2(x,\phi_\frkg\circ(\omega^\sharp)^{-1}(\xi)),\eta\rangle.
$$
Let $\xi=\omega^\sharp(y)$ and $\eta=\omega^\sharp(z)$, we have
\begin{eqnarray*}
\langle(\omega^\sharp)^{-1}\phi_\frkg^*l_2(x,\xi),\eta\rangle&=&-\langle\phi_\frkg^*\rho_{\phi_\frkg}^*(x)(\omega^\sharp(y)),z\rangle
=\langle\omega^\sharp(y),\rho_{\phi_\frkg}(x)(\phi_\frkg(z))\rangle=\omega(y,x\star
\phi_\frkg(z)),\\
\langle
l_2(x,\phi_\frkg\circ(\omega^\sharp)^{-1}(\xi)),\eta\rangle&=&\langle[x,\phi_\frkg(y)]_\frkg,\omega^\sharp(z)\rangle=\omega(z,[x,\phi_\frkg(y)]_\frkg).
\end{eqnarray*}
Since $\phi_\frkg^2=\Id$, we have
$$
\omega(y,x\star \phi_\frkg(z))=-\omega(\phi_\frkg^2(x)\star
\phi_\frkg(z),\phi_\frkg^2(y))=\omega(\phi_\frkg^2(z),
[\phi_\frkg^2(x),\phi_\frkg(y)]_\frkg)=\omega(z,[x,\phi_\frkg(y)]_\frkg),
$$
which implies that \eqref{eqn:confinal2} holds.

At last, let $\xi=\omega^\sharp(x)$ and $\eta=\omega^\sharp(y)$, the
equality \eqref{eqn:confinal3} is equivalent to
$$
\langle l_2(\phi_\frkg(x),\omega^\sharp(y)),z\rangle=\langle
l_2(\omega^\sharp(x),\phi_\frkg(y)),z\rangle.
$$
We have
\begin{eqnarray*}
\langle
l_2(\phi_\frkg(x),\omega^\sharp(y)),z\rangle&=&-\langle\omega^\sharp(y),\phi_\frkg(x)\star
z\rangle=\omega(\phi_\frkg(x)\star z,y)=-\omega(\phi_\frkg(z),[\phi_\frkg(x),\phi_\frkg(y)]_\frkg),\\
\langle
l_2(\omega^\sharp(x),\phi_\frkg(y)),z\rangle&=&\langle\omega^\sharp(x),\phi_\frkg(y)\star
z\rangle=\omega(x,\phi_\frkg(y)\star
z)=\omega(\phi_\frkg(z),[\phi_\frkg(y),\phi_\frkg(x)]_\frkg).
\end{eqnarray*}
Thus, \eqref{eqn:confinal3} holds. The proof is completed.\qed

\end{document}